\documentstyle[12pt]{article}

\newcommand{\ie}{{\em i.e.}}
\newcommand{\eg}{{\em e.g.}}
\newcommand{\R}{I\!\!R}
\newcommand{\DD}{{\cal D}}
\newcommand{\OO}{{\cal O}}

\newcommand{\lhs}{{\em lhs }}
\newcommand{\rhs}{{\em rhs }}
\newcommand{\eps}{\varepsilon}

\begin{document}

\pagestyle{plain}
\pagenumbering{arabic}

\author{P.~Exner and P.~\v Seba}

\title{Resonance statistics in a microwave cavity with a thin antenna}
\date{Nuclear Physics Institute, Academy of Sciences, \\
25068 \v Re\v z near Prague, \\ and
Doppler Institute, Czech Technical University, \\
B\v rehov\'a 7, 11519 Prague, Czech Republic \\ \rm exner@ujf.cas.cz,
seba@vtx.cz} 


\maketitle

\begin{abstract}
We propose a model for scattering in a flat resonator with a thin
antenna. The results are applied to rectangular microwave cavities.
We compute the resonance spacing distribution and show that it
agrees well with experimental data provided the antenna radius is
much smaller than wavelengths of the resonance wavefunctions.
\end{abstract}

\vspace{10mm}

\noindent
The aim of the present letter is twofold. First of all, we want to
present a method to treat scattering in systems consisting of
components of different dimensionality. To illustrate how it works,
we are going then to apply it to the case of a flat microwave cavity
to which a thin antenna is attached. We compute the resonance spacing
distribution in this system and show that it reproduces experimental
data --- within certain range of energies but without introducing any
free parameters.

Many recent studies involve an analysis, both theoretical and
experimental, of spectral and transport properties of systems with a
complicated geometry: let us recall various microwave resonators
\cite{AGH,GHL,KSP,Sri,SK} or conductance fluctuations in quantum dots
--- see \cite{JSA,PEI,BKW,FGH} and references therein.
Although physically different such systems share many common
properties, in particular, chaotic spectral behavior for certain
``resonator" shapes. For the sake of definiteness, we shall speak
here only about the electromagnetic case considering the experimental
setup of the work mentioned above, \ie, a flat microwave cavity which
may be regarded as two--dimensional coupled to one or more antennas
that supply power compensating radiative loss.

Already the very first experiments \cite{SSt} revealed a
discrepancy with theoretical predictions: the expected Poissonian
eigenvalue--spacing distribution for integrable rectangular
resonators was found to be distorted near the origin.  This effect
was later explained \cite{HLS} as consequence of the perturbation due
to the measuring antenna.

A complementary --- and more general --- point of view is to regard the
resonator with attached antennas as a scattering system and to study its
transport properties; this approach is also natural in view of the
mentioned analogy with conductance of quantum dots to which quantum
wire leads are attached. Solving the scattering problem in the
described geometry is not easy and one naturally looks for possible
model simplifications. Since the antennas are thin one can try to
describe them as lines coupled to the planar resonator in which the
field obeys the Laplace equation with Dirichlet boundary conditions.

At a glance the different dimensionality of the configuration space
parts adds complications to solution of the corresponding wave
equation. However, a standard technique based on self--adjoint
extensions \cite{ES1} allows us to reduce the task to matching 
solutions in different parts. We shall describe it below restricting
ourselves to simplest case with one antenna; a more general situation
will be discussed in a forthcoming paper \cite{ES2}. 

Let us begin with a halfline attached to a plane; we want to couple
the corresponding free Hamiltonians in a self--adjoint way. In
physical terms this requirement means the current conservation. It
will be ensured by boundary conditions (\ref{bc}) given below which
are local, and therefore also applicable when the halfline is
replaced by a line segment and/or the plane by any nonempty planar
region.  The halfline--plus--plane system corresponds to the Hilbert
space $\,L^2(\R_-)\oplus L^2(\R^2)\,$, \ie, the field is described by
pairs $\,\phi:= {\phi_1\choose\Phi_2}\,$ of square integrable
functions on the two parts of the configuration space. For the sake
of simplicity, and with the quantum mechanical analogy in mind, we
shall speak about wavefunctions.

The system has a rotational symmetry, so the partial--wave
decomposition may be employed. In particular, the s--wave component
$\,\Phi_2(r,\varphi)= (2\pi)^{-1/2}\phi_2(r)\,$ is independent of the
azimuthal angle.  The Hamiltonian acts on the wavefunction components
as the Laplacian, 
   \begin{equation} \label{Hamiltonian} 
H\phi\,=\, {-\phi''_1 \choose -\Delta\Phi_2}\,.
   \end{equation} 
The self--adjointness means a proper matching of the components at the
junction. A detailed discussion of this problem involving a full
classification of admissible boundary conditions was presented in
\cite{ES1}. They can be expressed in terms of generalized boundary
values for wavefunctions in the plane \cite{BG}
   \begin{eqnarray}
L_0(\Phi) &\!:=\!& \lim_{r\to 0}\,{\Phi(\vec x)\over\ln r}\,,
\nonumber \\ 
L_1(\Phi) &\!:=\!& \lim_{r\to 0}\,\left\lbrack\,
\Phi(\vec x)- L_0(\Phi) \ln r\, \right\rbrack \,. \label{generalized bv} 
   \end{eqnarray} 
Typical boundary conditions then read 
   \begin{eqnarray}
\phi'_1(0-) &\!=\!& A\phi_1(0-)+ BL_0(\Phi_2)\,, \nonumber \\ 
L_1(\Phi_2) &\!=\!& C\phi_1(0-)+ DL_0(\Phi_2)\;; \label{bc}
   \end{eqnarray}
there are additional lower--dimensional families corresponding to the
cases where (\ref{bc}) becomes singular \cite{ES1}. Let us remark
that using the appropriate regularized boundary values, one can
couple in a similar way a halfline to $\,\R^3$, and more generally,
two regions whose dimensions do not differ by more than two.

The coefficients $\,A,B,C,D\,$ were in the last named paper expressed
in terms of elements of the unitary matrix relating the deficiency
subspaces. In this way they depend on four real parameters; the
number reduces to three if we demand the Hamiltonian to be
time--reversal invariant. A drawback of these parametrizations is
that they do not show explicitly the ranges for the values of the
coefficients.

An alternative approach has been advocated recently in \cite{Ki}. It
is based on expressing the wavefunction in the plane as
$\,\Phi_2=\alpha G_0(\cdot,0;\sqrt{\lambda_0})+ u(\cdot)\,$, where
$\,G_0\,$ is the free Green's function for some energy value
$\lambda_0$ which does not belong to the spectrum of $H$ and $\,u\,$
is a function regular at the connection point $\,\vec x=0\,$. The
boundary conditions then tie $\,\alpha\,$ and $\,u(0)\,$ with
$\,\phi_1(0-),\, \phi'_1(0-)\,$. In general, one can obtain in this
way the same set of admissible Hamiltonians as above, however, in the
mentioned paper a particular subset corresponding to $\,A=0\,$ was
discussed and used.  

A more straightforward way is to compute the boundary form to
$\,H_0^*\,$, the adjoint to $\,H_0\,$ obtained by restriction of a
``decoupled" operator, say, the one with $\,B=C=0\,$, to functions
which vanish at the vicinity of the junction \cite{ES1}. Since the
action of $\,H_0^*\,$ is given by the same differential expression
(\ref{Hamiltonian}) we find after a simple integration by parts
   \begin{eqnarray}
\lefteqn{ 
(\phi, H_0^*\psi)- (H_0^*\phi,\psi) \,=\, \bar\phi'_1(0) \psi_1(0)-
\bar\phi_1(0) \psi'_1(0) }  \nonumber \\ &&  
\phantom{AAAA} +\lim_{\eps\to 0+}\, \eps \left(\bar\phi_2(\eps)
\psi'_1(\eps)-  \bar\phi'_2(\eps) \psi_2(\eps) \right)\,. \phantom{AA}
\label{bform}
   \end{eqnarray}
We know from \cite{ES1} that only the s--wave component in the plane
can be coupled nontrivially, hence we may suppose that $\,\Phi_2(\vec
x)\,$ is independent of the azimuthal angle and
$$
\phi_2(\eps)\,=\, \sqrt{2\pi}\,\left\lbrack L_0(\Phi_2)\ln\eps+
L_1(\Phi_2) +\OO(\eps)\, \right\rbrack\,.
$$
Using this asymptotic behavior we can rewrite the the last term on
the \rhs of (\ref{bform}) as 
$$
2\pi \left\lbrack L_1(\Phi_2) L_0(\Psi_2)-
L_0(\Phi_2) L_1(\Psi_2) \right\rbrack\,;
$$
it is then clear that the boundary form is zero under the conditions
(\ref{bc}) with 
   \begin{equation} \label{parameter range}
A,\,D\in\R \qquad {\rm and} \qquad B=2\pi\bar C\,.
   \end{equation}
If the Hamiltonian should be time--reversal invariant, the boundary
conditions must not change when passing to complex conjugated
wavefunctions; this happens if the off-diagonal coefficients $B,C$ are
real. 

To proceed further we have specify the coefficients in (\ref{bc}).
The choice needs a physical motivation. Since we are interested
in transport through such junctions, we shall use a comparison of
low--energy scattering properties with those of a more realistic
antenna. The reflection and transmission amplitudes for a plane wave
$\,e^{ikx}\,$ approaching the junction on the halfline were computed
in \cite{ES1}: we have
   \begin{equation} \label{rt point}
r(k)\,=\,-\,{\DD_-\over\DD_+}\,,\qquad t(k)\,=\,{2iCk\over\DD_+}\,,
   \end{equation}
where
$$
\DD_{\pm}\,:=\,(A\pm ik)\left\lbrack\, 1+{2i\over\pi} \left(\gamma-D
+\ln{k\over 2} \right)\right\rbrack +{2i\over\pi}\, BC
$$
and $\,\gamma= 0.5772\dots\,$ is the Euler's constant.

Consider now a semiinfinite cylindrical tube of radius $\,a\,$
attached to the plane; we suppose the latter has the central circular
area of radius $\,a\,$ removed so that the whole surface is
topologically equivalent to an infinite cylinder. The scattering on
such a surface is certainly affected by the sharp edge at the
interface of the two parts but we disregard this fact.

We have a rotational symmetry again, so each partial wave may be
treated separately. Suppose that the ``longitudinal" component of the
incident wavefunction is a plane wave of momentum $\,k\,$. For an
orbital quantum number $\,\ell\,$ one has to match smoothly the
corresponding solutions
$$
f(x)\,:=\, \left\lbrace \begin{array}{ccc} e^{ikx}+r_a e^{-ikx} &
\quad \dots \quad & x\le 0 \\ \\ \sqrt{\pi kr\over 2}\,t_a
H_\ell^{(1)}(kr) & \quad \dots \quad & r\ge a \end{array} \right.
$$
We abuse here notation on the \lhs and employ $\,x\,$ as a common
denomination of the longitudinal variable on the cylinder and the
radial variable in the plane. The matching conditions are solved
easily giving
   \begin{equation} \label{rt tube}
r_a(k)\,=\, -\,{\DD^a_-\over\DD^a_+}\,, \qquad 
t_a(k)\,=\, 4i\, \sqrt{2ka\over\pi}\, \left(\DD^a_+\right)^{-1}
   \end{equation}
with
$$
\DD^a_{\pm}\,:=\, (1\pm 2ika)H_\ell^{(1)}(ka)+ 2ka
\left(H_\ell^{(1)}\right)'(ka) \;;
$$
the Wronskian relation $\,W(J_\nu(z),Y_\nu(z))= 2/\pi z\,$ implies
that the unitarity requirement, $\,|r_a(k)|^2\!+ |t_a(k)|^2=1\,$, is
satisfied. 

It follows from the asymptotic properties of Bessel functions
\cite{AS} that $\,|t_a(k)|^2 \sim (ka)^{2\ell-1}\,$
holds for $\,\ell\ge 1\,$, so the transmission probability vanishes
fast as $\,k\to 0\,$ for higher partial waves. On the other hand,
$$
H_0^{(1)}(z)\,=\, 1+\,{2i\over\pi} \left(\gamma+\, \ln{ka\over 2}
\right)+ \OO(z^2\ln z)\,.
$$
Substituting into (\ref{rt tube}) we find that these amplitudes have
for $\,ka\ll 1\,$ the same leading behavior as (\ref{rt point})
provided we put
   \begin{equation} \label{identification}
A:=\, {1\over 2a}\,, \qquad D:= -\ln a\,, \qquad BC:=\, {1\over a}\;;
   \end{equation}
owing to (\ref{parameter range}) the last relation is in the case of
``real" boundary conditions satisfied if
   \begin{equation} \label{offdiagonal identification}
B\,=\, 2\pi C\,=\, \sqrt{2\pi\over a}\;.
   \end{equation}
In this way we have found values of the coefficients for which the
coupling (\ref{bc}) is able to model the scattering behavior of a
real antenna as long as the wavelengths involved are much larger than
the antenna radius. A similar effect has been noticed recently for
point interactions in dimension two or three \cite{ES3}: long enough
waves feel only the size of an obstacle or a drain. It is also worth
noting that the coefficient $\,A\,$ is always nonzero by
(\ref{identification}), so the coupling used in Ref.\cite{Ki} does
not belong to this class.

Now we are ready to describe our model. We consider a
resonator in the form of a planar region $\,M\,$ to which an antenna 
is attached at a point $\,x_0\in M\,$; we suppose
that at the junctions the wavefunctions are coupled by boundary
conditions (\ref{bc}) with the parameter values
(\ref{identification}), (\ref{offdiagonal identification}), and that
the resonator has hard walls which we model by the Dirichlet
condition, $\,u(\vec x)=0\,$ at the boundary of $\,M\,$.

We have to match the plane wave combination $\,e^{ikx}+r\,e^{-ikx}\,$
on the antenna halfline $\,\R_-\,$ with the internal solution
$\,u(\vec x):=bG(\vec x,\vec x_1;k)\,$, where $\,G(\cdot,\cdot;k)\,$
is the Green's function of the Dirichlet Laplacian; we assume, of
course, that $\,k^2$ equals to no eigenvalue of this operator. To
make use of (\ref{bc}) we have to know the generalized boundary
values $\,L_i:= L_i(u;\vec x_0)\,$; they are
   \begin{equation} \label{boundary values}
L_0=\,-\, {b\over 2\pi}\,, \qquad L_1= b\xi(\vec x_0;k)\,,
   \end{equation}
where 
   \begin{equation} \label{xi}
\xi(\vec x_0;k):=\, \lim_{\vec x\to \vec x_0} \left\lbrack\, G(\vec
x,\vec x_0;k)+\, {\ln|\vec x-\vec x_0|\over 2\pi}\, \right\rbrack\,.
   \end{equation}
Matching then the solutions we find
   \begin{equation} \label{rr1}
r(k)\,=\, -\, {\pi Z(k)(1-2ika) -1 \over \pi Z(k)(1+2ika) -1}
   \end{equation}
with
   \begin{equation} \label{Z}
Z(k)\,:=\, \xi(\vec x_0;k)-\, {\ln a\over 2\pi}
   \end{equation}
In order to make use of these formulae we have to know the last named
quantity. If the region $\,M\,$ is compact, the corresponding
Dirichlet Laplacian has a purely discrete spectrum.  We denote by
$\,\lambda_n\,$ amd $\,\phi_n\,$, respectively, the corresponding
eigenvalues and eigenfunctions; without loss of generality the latter
can be chosen real. This allows us to express the Green's function,
   \begin{equation} \label{g}
G(\vec x_1,\vec x_2;k)\,=\, \sum_{n=1}^{\infty}\, {\phi_n(\vec x_1)
\phi_n(\vec x_2) \over \lambda_n-k^2}\,.
   \end{equation}
It diverges, of course, as $\,\vec x_1\to \vec x_2\,$; we need to
know the regularized value (\ref{xi}). A semiclassical argument
\cite[Sec.XIII.15]{RS} makes it possible to asses the rate of the
divergence: we have $\,\lambda_n\approx 4\pi n|M|^{-1}\,$ and
$\,\langle |\phi_n(\vec x)|^2\rangle \approx |M|^{-1}\,$ as $\,n\to
\infty\,$, where $\,|M|\,$ is the area of the resonator. This
inspires us to employ the identity
   \begin{eqnarray*}
\lefteqn{
G(\vec x_0+\sqrt\eps\vec n,\vec x_0;k)+\, {\ln
\sqrt\eps\over 2\pi}\,=\,} \\ && 
\sum_{n=1}^{\infty}\, \left\lbrace\, {\phi_n(\vec
x_0+\sqrt\eps\vec n)  
\phi_n(\vec x_0) \over \lambda_n-k^2}
\,-\, {(1-\eps)^n\over 4\pi n} \, \right\rbrace\;,
   \end{eqnarray*}
where $\,\vec n\,$ is an arbitrary unit vector. The two series
diverge as $\,\eps\to 0\,$ but their difference remains finite: we
have  
   \begin{equation} \label {xi explicit}
\xi(\vec x_0;k)\,=\, \sum_{n=1}^{\infty}\, \left\lbrace\,
{|\phi_n(\vec x_0)|^2 \over \lambda_n-k^2}\,-\, {1\over 
4\pi n} \, \right\rbrace\;.
   \end{equation}
Let us stress that the exchange of the limit and summation should be
taken with a caution, since the series is not uniformly convergent.
Replacing, \eg, $\,\sqrt\eps\,$ by $\,c\sqrt\eps\,$ we get an
additive factor. However, the correct $\,\xi(\vec x_0;k)\,$ should be
independent of the boundary at large negative energies, 
$$
\lim_{\kappa\to\infty} \left\lbrack \xi(\vec x_0;i\kappa) -\, {1\over
2\pi}\, \left( \ln{\kappa\over 2}\,+\gamma \right) \right\rbrack
\,=\,0 \,,
$$
by Ref.\cite{ES1}. For a specific $\,M\,$ as the one considered
below, it is straightforward to check that this is true for (\ref{xi
explicit}). 

We are especially interested in the situation where $\,M\,$ is a
rectangle $\,[0,c_1]\times [0,c_2]\,$, for which we have 
   \begin{eqnarray} 
\phi_{nm}(x,y) \!&=&\! \frac{2}{\sqrt{c_1 c_2}}\, \sin(n\frac{\pi}{c_1}x) 
\sin(m\frac{\pi}{c_2}y)\,, \label{rectangle ef} \\ \nonumber \\ 
\lambda_{nm} \!&=&\! \frac{n^2\pi^2}{c_1^2}\!
+\frac{m^2\pi^2}{c_2^2}\;; \label{rectangle ev}
   \end{eqnarray}
the formulae (\ref{rr1}) and (\ref{xi explicit})--(\ref{rectangle
ev}) yield a complete solution of the scattering problem in this
setting. 

In particular, we want to find the resonance spacing distribution
which can be compared with the result found experimentally in
\cite{HLS}; we refer to this paper for a more detailed description of
the experimental arrangement. By (\ref{rr1}) the resonances are given
by complex zeros of the denominator, \ie, by solutions of the
algebraic equation 
\begin{equation}
\xi(\vec x_0,k)\,=\,{\ln(a)\over 2\pi}\,+\,{1\over\pi(1+ika)}
\end{equation}
where $\,a\,$ is the radius of the antenna opening. We have solved
this equation numerically and evaluated the resonance
spacing distribution. In order to mimick the experimental setting 
we have evaluated the resonance spaing distribution for several 
rectangular resonators with $\,c_1\,$ and $\,c_2\,$ ranging 
from 20 to 50 cm. The coupling point 
$\,\vec x_0\,$ has been chosen randomly for each billiard and 
the diameter of the antena was equal to $\,1\, {\rm mm}$.
The obtained data has been combined in order to enlarge the 
statistics. Since the present
model was derived under the assumption $\,ka\ll 1\,$ we used only the
data corresponding to frequencies below 10 GHz for which
$\,ka<0.1\,$. In addition, it is known that about 7\% of resonances
are overlooked in the experiment; to take the missing levels into
account we have deleted the same amount of randomly chosen resonances
from the family under consideration. 


The result is plotted on the Figure 1. The agreement between the 
model calculations and the experiment data is convincing. We recall
that our argument involves no free parameters because the ``coupling
strength" between resonator and the antenna is fully determined by
the size of the latter. On the other hand, the long--wave condition
relative to the antenna radius, $\,ka \ll 1\,$, is important: the
agreement worsens if resonances above 10 GHz are taken into account.

In conclusion, we have described a method to treat the scattering
problem in systems with configuration space composed of parts of a
different dimensionality. It has been applied to a flat microwave
resonator with a thin antenna; the calculated resonance spacing
distribution agrees in the long--wave regime with the experimental
data.

\subsection*{Acknowledgments} 

The research has been partially supported by the Grant AS CR
No.148409.

\subsection*{Figure captions} 

\begin{description}
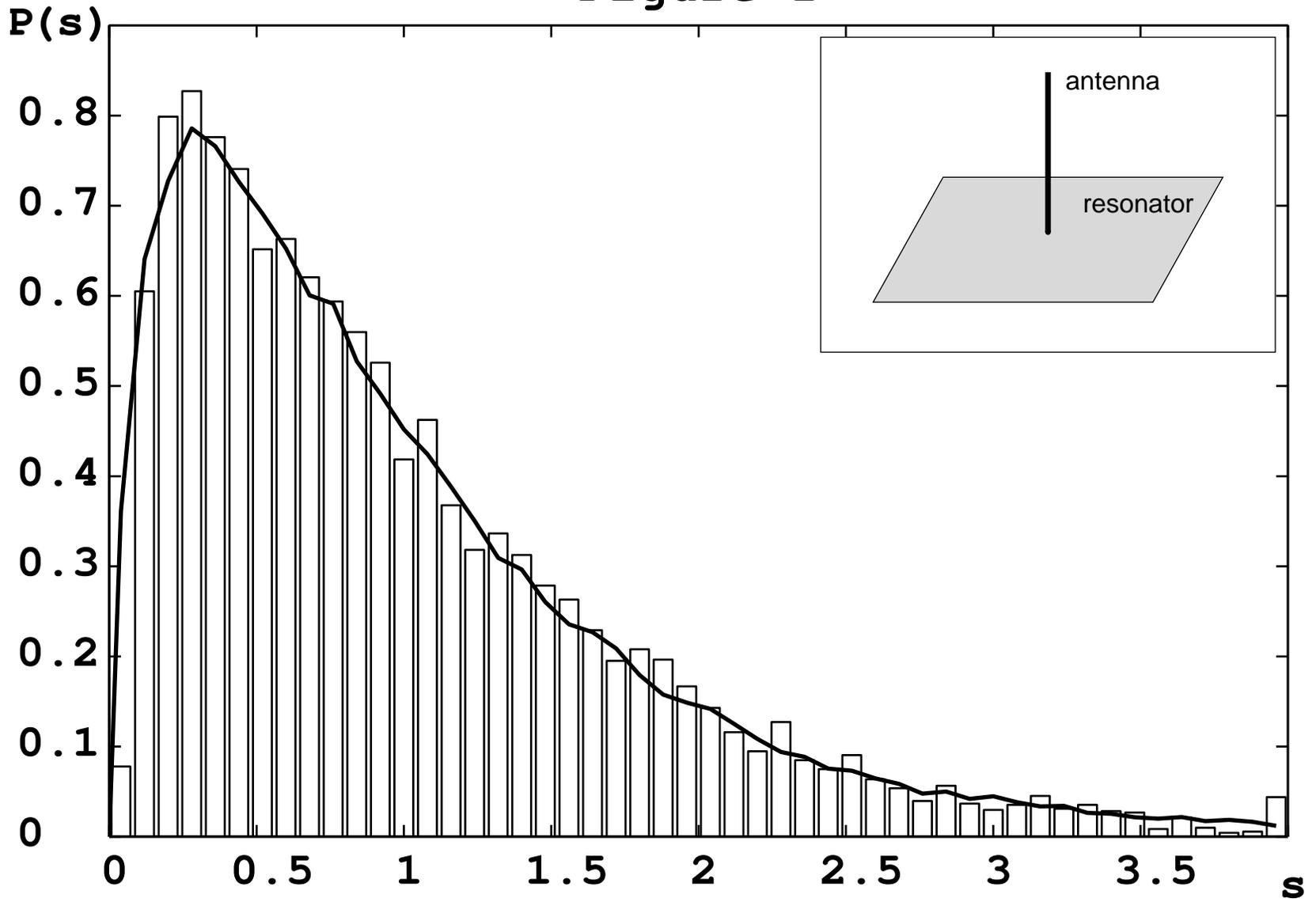
\item{\bf Figure 1.}\ The numerically evaluated resonance spacing 
distribution for the rectangular resonator (full line) in comparison
with the experimental data taken from \cite{HLS} (bins). In the inset
we sketch the geometric arrangement of the experiment.
\end{description}

\end{document}